\documentclass[
    ,final            
  ]
  {aipproc}

\layoutstyle{6x9}

\newcommand{\etal}{et al.}

\newcommand{\gtsim}{\raisebox{-1mm}{$\stackrel{>}{\sim}$}}

\begin{document}

\title{The XMM-Newton Slew Survey: Towards The Whole X-ray Sky and the
  Rarest X-ray Events}


\author{A.M. Read}{
  address={University of Leicester, Leicester, LE1 7RH, UK}
}

\author{R.D. Saxton}{
  address={XMM SOC, ESAC, Villanuevade la Canada, Madrid, Spain}
}

\author{P. Esquej}{
  address={University of Leicester, Leicester, LE1 7RH, UK}
}

\author{R.S. Warwick}{
  address={University of Leicester, Leicester, LE1 7RH, UK}
}

\begin{abstract}

  The data collected by XMM-Newton as it slews between pointings
  currently cover almost half the entire sky, and many familiar features 
  and new sources are visible. The soft-band sensitivity limit
  of the Slew is close to that of the RASS, and a large-area Slew-RASS
  comparison now provides the best opportunity for discovering
  extremely rare high-variability objects.

\end{abstract}

\maketitle


\section{Towards The Whole X-ray Sky}

\begin{figure}[b]
  \includegraphics[bb=153 53 476 739,height=.65\textheight,angle=270]{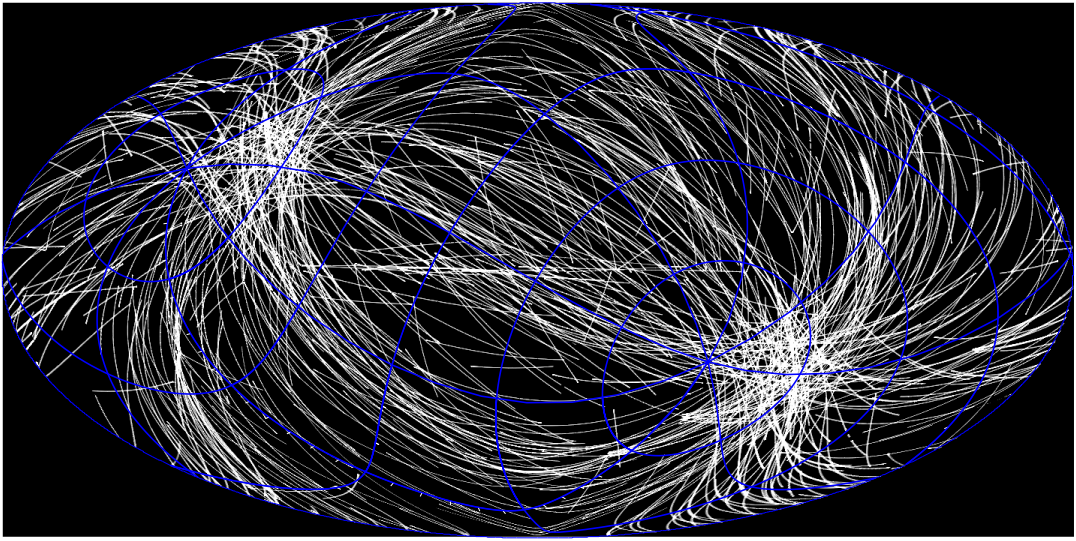}
  \caption{The full XMM-Newton slew exposure map (third catalogue update; June
    2009), in a Galactic Aitoff co-ordinate projection. It covers
    $\approx41$\% of the sky, with exposures ranging from just a few
    seconds at the slew path edges, up to close to a minute in the overlap
    regions near the Ecliptic poles.}
\end{figure}

The publicly available XMM-Newton slew data cover to date around 41\%
of the sky (see Fig.\,1). The soft band (0.2$-$2 keV) sensitivity
limit of the slews (6$\times10^{-13}$\,ergs cm$^{-2}$ s$^{-1}$) is
close to that of the ROSAT All-Sky Survey (RASS), and in the medium
(2$-$12 keV) band, the slew data goes significantly deeper
(4$\times10^{-12}$\,ergs cm$^{-2}$ s$^{-1}$) than all other previous
large area surveys. The current full catalogue (the third update,
released in June 2009) contains 19,720 detections. For details on 
the construction and characteristics of the XMM-Newton slew survey
catalogue, see Saxton et al. (2008), and for details of the initial
science results from the slew survey, see Read et al. (2006).

\section{The Rarest X-ray Events}

The near real-time comparison of XMM-Newton slew data with RASS data
is giving, for the first time, the opportunity of finding all manner
of high-variability X-ray objects, e.g. tidal disruption candidates
(Esquej \etal\ 2007, and this conference), AGN, blazars, and also
Galactic sources such as novae, flare stars, cataclysmic variables,
and eclipsing X-ray binaries. It is only with a large-area survey,
such as the XMM-Newton Slew Survey, that such rare events have a
chance of being caught. 
Two such high-variability sources we have classified as new classical
novae. XMMSL1~J070542.7-381442 (Read \etal\ 2008), also known as V598
Pup, was seen in the slew to be $\sim$750 times brighter than the RASS
upper limit. Our Magellan optical spectrum identified the source as an
auroral phase nova, and the X-ray data indicates that the nova was in
a super-soft state (see Fig.\,2), and at a distance of
$\sim$3\,kpc. Analysis of archival optical ASAS data showed an
incredibly bright (and at the time un-noticed) source, rising from
m$_{V}\gtsim$14 to m$_{V}$$\approx$4.1 over three days, making it
one of the very brightest novae this decade.  A second source,
XMMSL1~J060636.2-694933 (Read \etal\ 2009), was seen in the slew to be
$\gtsim$500 times brighter than in the RASS, and XMM-Newton, Swift,
Magellan and ASAS data were able to classify it as again an auroral
phase, very fast nova in the super-soft state (Fig.\,2), though at a
much further distance and likely situated in the LMC.
With discoveries such as these, it is clear that XMM-Newton slew data
are continuing to offer a powerful opportunity to find new X-ray
transient objects.

\begin{figure}
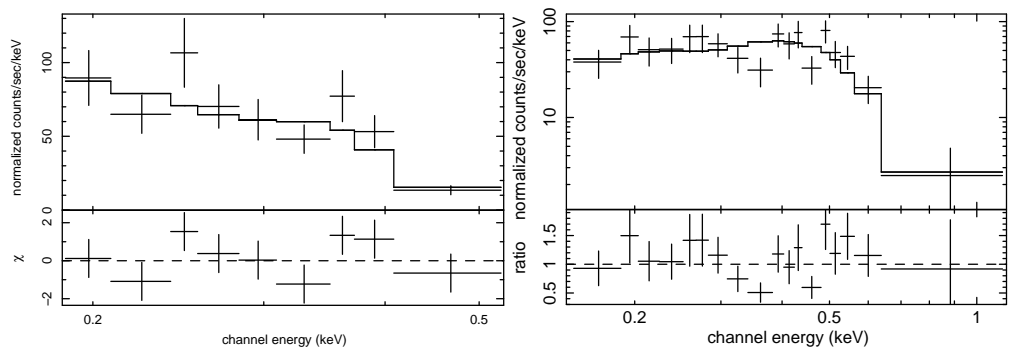


  \includegraphics[bb=112 22 540 700,height=.3\textheight,angle=270]{0705_slewspec}
  \includegraphics[bb=112 20 571 700,height=.3\textheight,angle=270]{0606_slewspec}
  \caption{XMM-Newton slew EPIC-pn spectra of (left) XMMSL1~J070542.7-38144 and (right)
    XMMSL1~J060636.2-694933. Both are fitted with very low temperature
    black-body models.}
\end{figure}


\end{document}